\documentclass[fleqn,10pt]{wlscirep}
\usepackage[utf8]{inputenc}
\usepackage{lineno}
\usepackage[T1]{fontenc}
\usepackage{subcaption}
\title{Collaboration in the Time of COVID: A Scientometric Analysis of Multidisciplinary SARS-CoV-2 Research}

\author[1,2,*]{Eoghan Cunningham}
\author[1,2]{Barry Smyth}
\author[1,2]{Derek Greene}
\affil[1]{Insight SFI Research Centre for Data Analytics, University College Dublin, Ireland}
\affil[*]{eoghan.cunningham@ucdconnect.ie}
\begin{abstract}
The novel coronavirus SARS-CoV-2 and the COVID-19 illness it causes have inspired unprecedented levels of multidisciplinary research in an effort to address a generational public health challenge. In this work we conduct a scientometric analysis of COVID-19 research, paying particular attention to the nature of collaboration that this pandemic has fostered among different disciplines. Increased multidisciplinary collaboration has been shown to produce greater scientific impact, albeit with higher co-ordination costs. As such, we consider a collection of over 166,000 COVID-19-related articles to assess the scale and diversity of collaboration in COVID-19 research, which we compare to non-COVID-19 controls before and during the pandemic. We show that COVID-19 research teams are not only significantly smaller than their non-COVID-19 counterparts, but they are also more diverse. Furthermore, we find that COVID-19 research has increased the multidisciplinarity of authors across most scientific fields of study, indicating that COVID-19 has helped to remove some of the barriers that usually exist between disparate disciplines. Finally, we highlight a number of interesting areas of multidisciplinary research during COVID-19, and propose methodologies for visualising the nature of multidisciplinary collaboration, which may have application beyond this pandemic. 
\end{abstract}

\begin{document}
\flushbottom
\maketitle
\thispagestyle{empty}

%\noindent Please note: Abbreviations should be introduced at the first mention in the main text – no abbreviations lists. Suggested structure of main text (not enforced) is provided below.

\section*{Introduction}
The scientific response to the SARS-CoV-2 pandemic has been unprecedented with researchers from several surprising fields --- e.g. artificial intelligence \cite{nguyen_2021}, economics \cite{nicola_2020}, and particle physics \cite{lustig_2020} --- contributing to solving the many and varied clinical and societal challenges arising from the pandemic. As a result, by January 2021, The Allen Institute for AI\cite{allenai} and the World Health Organisation\cite{who} had identified over 166,000 research papers relating to SARS-CoV-2 and the COVID-19 illness it causes, highlighting an unprecedented period of scientific productivity. In this study we analyse this body of work to better understand the scale and nature of the collaboration and fields of study that have defined this research. 

% As of January 2021, The Allen Institute for AI\cite{allenai} and the World Health Organisation\cite{who} had identified over 160,000 research papers relating to the novel coronavirus, SARS-COV2, and the COVID-19 illness it causes. To a largely unprecedented degree, the scientific community has turned its attention to this pandemic. Researchers from fields far beyond those of epidemiology and virology, such as particle physics \cite{lustig_2020}, artificial intelligence \cite{nguyen_2021}, and economics \cite{nicola_2020}, have attempted to tackle the clinical, pharmaceutical, and societal challenges arising from the pandemic. The resulting research is distinguished by its productivity and diversity. 

The benefits of collaboration during scientific research are well documented and widely accepted and in recent years there has been steady growth in research team size across all scientific disciplines \cite{leahey_2016,youngblood_2018}, which has been shown to correlate positively with research impact \cite{lariviere_2014,porter_2008}. Moreover, multidisciplinary science, which brings together researchers from many disparate subject areas has been shown to be among the most successful scientific endeavours \cite{okamura_2019,lariviere_2015}. Indeed, multidisciplinary research has been highlighted as a key enabler when it comes to addressing some of the most complex challenges facing the world today \cite{leahey_2016}. Not surprisingly then, there have been numerous attempts to encourage and promote collaboration and cooperation in the fight against COVID-19: the World Health Organisation maintains a COVID-19 global research database; scientific journals have published explicit calls for teamwork and cooperation \cite{chakraborty_2020,nature_call_for_papers}; in many cases COVID-19 related research has been made freely available to the public and the scientific community; comprehensive datasets have been created and shared; and reports from the International Chamber of Commerce (ICC) and the Organisation for Economic Co-operation and Development (OECD) have argued for international and multidisciplinary collaboration in the response to the pandemic.

Although early studies have found that the pandemic has generated a significant degree of novel collaboration \cite{hook_2020,liu_2021} other research has suggested that COVID-19 research have been less internationally collaborative than expected, compared with recent research from the years immediately prior to the pandemic \cite{fry_2020,hook_2020}. There is also some evidence that COVID-19 teams have been smaller than their pre-2020 counterparts \cite{cai_2021,fry_2020}. Thus, despite calls for greater collaboration, the evidence points to less collaboration in COVID-19 related research, perhaps because of the startup and coordination costs associated with multidisciplinary research \cite{cai_2021,fry_2020,hook_2020} combined with the urgency of the pandemic response that has been needed.

% suggest that, due to the urgency of COVID-19 research and the greater coordination costs of international and multidisciplinary collaboration, research related to the pandemic has been less collaborative than would be desired or expected for a challenge of this magnitude \cite{cai_2021,fry_2020,hook_2020}. 

% . To do this we analyse three groups research articles: 

% \begin{enumerate}
%     \item \emph{2020-COVID-related} research: COVID-related articles published during the pandemic (2020);
%     \item \emph{pre-2020} research: \emph{non-COVID related} articles published \emph{before} the pandemic, that is prior to 2020;
%     \item \emph{2020-non-COVID} research: additional non-COVID related articles published \emph{during} the pandemic period.
% \end{enumerate}
% (i) \emph{2020-COVID-related} research, which refers to COVID-related articles published during the pandemic (2020); (ii) \emph{pre-2020} research, which refers to (\emph{non-COVID related}) articles published before the pandemic, that is prior to 2020; and (iii) \emph{2020-non-COVID} research, which refers to non-COVID related articles published during the pandemic period. 

In this study we evaluate the scale and nature of collaboration in COVID-19 research during 2020, using scientometric analysis techniques to analyse COVID and non-COVID publications before (non-COVID) and during (COVID and non-COVID) the pandemic. We determine the nature of collaboration in these datasets using three different collaboration measures: (i) the \emph{Collaboration Index} (CI) \cite{youngblood_2018}, to estimate the degree of collaboration in a body of research; (ii) \emph{author multidisciplinarity} to estimate the rate at which authors publish in different disciplines; and (iii) \emph{team multidisciplinarity} to estimate subject diversity across research teams. We find a lower CI for COVID-related research teams, despite an increasing CI trend for non-COVID work, before and during the pandemic, but COVID-related research is associated with higher author multidisciplinarity and more diverse research teams. This research can help us to better understand the nature of the research that has been conducted under pandemic conditions, which may be useful when it comes to coordinating similar large-scale initiatives in the future. Moreover, we develop a number of techniques for exploring the nature of collaborative research, which we believe will be of general interest to academics, research institutions, and funding agencies.

\section*{Methods}
In this section we describe our methods for evaluating scientific collaboration in COVID-19 research. We describe the data that we use throughout our analysis, and we outline three approaches used to evaluate collaboration activity.

\subsection*{Datasets}
The COVID-19 Open Research Dataset (CORD-19) \cite{wang_2020} comprises more than 400,000 scholarly articles, including over 150,000 with full text, all related to COVID-19, SARS-CoV-2, and similar coronaviruses. CORD-19 papers are sourced from PubMed, PubMed Central, bioRxiv, medRxiv, arXiv, and the World Health Organisation's COVID-19 database. We generate a set \emph{COVID-19-related} research by excluding articles dated prior to 2020 and the resulting dataset contains CORD-19 metadata for 166,356 research papers containing the terms “COVID”, “COVID-19”, “Coronavirus”, “Corona virus”, “2019-nCoV”, “SARS-CoV”, ”MERS-CoV”, “Severe Acute Respiratory Syndrome” or “Middle East Respiratory Syndrome”. We supplement this metadata with bibliographic information from the Microsoft Academic Graph (MAG) \cite{sinha_2015}. 

Notably, we use the MAG \emph{fields of study} (FoS) to categorise research papers. The MAG uses hierarchical topic modelling to identify and assign research topics to individual papers, each of which represents a specific field of study. To date, this approach has identified a hierarchy of over 700,000 topics within the Microsoft Academic Knowledge corpus. In our dataset of 166,356 COVID-19 research articles, the average paper is associated with 9 FoS from different levels in this hierarchy and in total, 65,427 unique fields are represented. To produce a more useful categorisation of articles, we first reduce the number of topics by replacing each field with its parent and then consider topics at two levels in the FoS hierarchy: (i) the 19 FoS at level 0, which we refer to as \textit{'disciplines’}, and (ii) the 292 FoS at level 1, which we refer to as \textit{‘sub-disciplines'}. In this way, each article is associated with a set of disciplines (e.g. 'Medicine', 'Physics', 'Engineering') and sub-disciplines (e.g. 'Virology', 'Particle Physics', 'Electronic Engineering'), which are identified by traversing the FoS hierarchy from the fields originally assigned to the paper. 

\begin{table}[h]
	\centering
	\begin{subtable}[h]{0.6\textwidth}
		\begin{tabular}{llll}
			\toprule
			{}                 & Authors   & Articles  & Fields  \\
			\midrule
			pre-2020           & 6,379,612 & 4,017,655 & 283,599 \\
			2020-non-COVID     & 3,200,107 & 1,205,434 & 196,409 \\
			2020-COVID-related & 627,205   & 166,356   & 65,427  \\
			\bottomrule
		\end{tabular}
		\caption{A summary of the numbers of authors, articles, and fields collected in the three datasets used in this study.}
	\end{subtable}
	\quad
	\begin{subtable}[h]{0.3\textwidth}
		\begin{tabular}{ll}
			\toprule
			Year & Articles  \\
			\midrule
			2016 & 954,174   \\
			2017 & 1,006,394 \\
			2018 & 987,666   \\
			2019 & 1,069,421 \\
			\midrule
			2020 & 1,371,190 \\
			\bottomrule
		\end{tabular}
		\caption{Articles per year across all datasets.}
	\end{subtable}
	\caption{\label{tab:dataser}Dataset summary: Note, all non-COVID research articles contain at least one author who published COVID-related research. Also, of the authors who published COVID-related research, we are able to collect non-COVID research (both pre-2020 and 2020-non-COVID) for 299,046 individuals.}
\end{table}

%BS. This table needs a bit of attention. It wasnt refed in main text so I have added that below. I think we also need to label and caption each table. I'm not sure I fully understand the table on the right. Does it related to total articles (across covid and non-covid?).

We further extend this dataset with any additional research published by the authors in the COVID-related dataset. Thus, for each author, we include MAG metadata from any available articles dated after 2015. The final dataset consists of metadata for 5,389,445 research papers, which we divide into three distinct groups as follows; see Table \ref{tab:dataser} with further detail provided in the supplementary materials that accompany this article (Supplementary Table 1-3).

\begin{enumerate}
    \item \emph{2020-COVID-related} research:  the 166,356 COVID-related articles published during the pandemic (2020);
    \item \emph{pre-2020} research: 4,017,655 \emph{non-COVID related} articles published \emph{before} the pandemic, that is during 2016--2019, inclusive;
    \item \emph{2020-non-COVID} research: 1,205,434 non-COVID related articles published \emph{during} the pandemic period and which are not in the CORD dataset.
\end{enumerate}

% A breakdown of the number of articles identified in each discipline is included in supplementary materials (Supplementary Table 1-3). 

% \emph{2020-COVID-related} research; \emph{2020-non-COVID} research (1,205,434 papers published in 2020, but not represented in the CORD-19 set); and \emph{pre-2020} research (4,017,655 papers published 2016-2019 inclusive). Thus the pre-2020 and 2020-non-COVID datsets are used as non-COVID baselines against which we evaluate COVID-related research. 

\subsection*{Collaboration Index}
The Annual Collaboration Index (CI) is defined, for a body of work, as the ratio of the number of authors of co-authored articles to the total number of co-authored articles \cite{youngblood_2018}. Since larger (more collaborative) teams have been shown to be more successful than smaller teams \cite{lariviere_2014,leahey_2016,klug_2014}, we can use CI to compare COVID-related research to non-COVID baselines. However, CI is sensitive to the total number of articles in the corpus. Therefore, to facilitate comparison across our COVID and non-COVID baselines we generate a CI distribution for each dataset by re-sampling 50,000 papers 1,000 times, without replacement, from each year, and we calculate the sample distribution for these CI values for each year in our dataset.

% ; see Figure \ref{fig:collaboration_index} reports the sample distribution of these CI values for each year in the dataset. We also include distributions for CI values for 2020-COVID-related and 2020-non-COVID samples. 

%BS: It is my understanding that we dont present results here.

\subsection*{Author Multidisciplinarity}
To evaluate the multidisciplinarity of individual authors, we consider the extent to which they publish across multiple disciplines, based on a network representation of their publications. An un-weighted bipartite network, populated by research fields and authors, links researchers to subjects (that is, based on the subjects of their publications). A projection of this network produces a dense graph of the 292 sub-disciplines at level 1 in the MAG FoS hierarchy, with two sub-disciplines/fields are linked if an author has published work in both. We refer to this projection as a \emph{field of study network}. In such a network, the edges between fields are weighted according to the number of authors publishing in both fields. Due to the large number of researchers, and the relatively small number of sub-disciplines, the resulting graph is almost fully connected. Thus, the edge weights are an important way to distinguish between edges. Using the MAG FoS hierarchy, we divide the network nodes into 19 overlapping ``communities'', based on their assignment to level 0 fields of study. This facilitates the characterization of the edges in the graph: an edge \emph{within} a community represents an author publishing in two sub-disciplines within the same parent discipline, while an edge \emph{between} communities represents an author publishing in two sub-disciplines from different parent disciplines. For example, if an author publishes research in 'Machine Learning' and 'Databases', the resulting edge is considered to be \textit{within} the community/discipline of 'Computer Science'. Conversely, if an author publishes in 'Machine Learning' and 'Radiography', the resulting edge is considered to be \textit{between} the 'Medicine' and 'Computer Science' communities. An edge between disciplines may represent either a single piece of interdisciplinary research or an author publishing separate pieces of research in two different disciplines. To evaluate the effect of COVID-19 on author multidisciplinarity, we produce a field of study network for each year in our dataset and calculate the proportion of the total edge weights that exist between communities. In the special case of 2020 we also report the odds ratio achieved when we compare 2019 with non-COVID research in 2020 (i.e., after we remove COVID-19 research from the graph).

% Figure \ref{fig:interdisciplinary_publication} reports the odds ratio effect size when the proportion of the edges that are \textit{between} communities is compared with the previous year. These scores are reported for each community. In the case of 2020 we also report the odds ratio achieved when we compare 2019 with non-COVID research in 2020 (i.e., after we remove COVID-19 research from the graph). 

%BS. It is my understanding that you don't present results in the Methods section. Therefore it might be better to shift the above para to the result section. I've commented it out for now.

\subsection*{Research Team Disciplinary Diversity}
In addition to author multidisciplinarity, we also consider the multidisciplinarity of the research teams, by calculating their \emph{disciplinary diversity}. To do this we compare the research backgrounds of different authors using publication vectors based on the proportions of a researcher’s work published across different fields \cite{feng_2020}. Specifically, we construct publication vectors for authors in our dataset using the 19 MAG disciplines. Thus, an author's publication vector is a 19-dimensional vector, with each value indicating the proportion of the author's research published in the corresponding domain. For example, an author who has 50 publications classified under 'Computer Science', 30 publications classified under 'Mathematics', and 20 publications classified under 'Biology' would have a publication vector with values \{0.5,0.3,0.2\} for the entries corresponding to these disciplines respectively, and zeros elsewhere. 
By using publication vectors to represent an individual’s research profile, we can quantify the disciplinary diversity of a research team using Equation \ref{equation:team_similarity} from \cite{feng_2020}.

\begin{equation}\label{equation:team_similarity}
	S_{team} := \frac{2}{|p|(|p| -1)}\sum_{(i,j)\in p}S_{ij}
\end{equation}

Note, in Equation \ref{equation:team_similarity} $|p|$ refers to the size of the research team and $S_{ij}$ is the cosine similarity of the publication vectors for authors $i$ and $j$. The team research similarity score for an article is a normalized sum of the pairwise cosine similarities for all authors of the article.

To evaluate research team disciplinary diversity, we compute the teams’ disciplinary similarity based on publication vectors from pre-2020 research, and we report $1 - S_{team}$ as the teams’ diversity. The year of the paper is excluded from the publication vector to avoid reducing team diversity with the common publication. As such, team disciplinary diversities for COVID-related research (and non-COVID research from 2020) are calculated from publication vectors which exclude work from 2020. We compare these scores with disciplinary diversity scores for research in 2019 when, similarly, the publication vectors exclude work from 2019 and 2020. As the potential for disciplinary diversity in research teams is limited by the number of team members, we compare diversity by team size. 

% Figure \ref{fig:research_team_diversity} illustrates the increase in mean disciplinary diversity for research teams of different sizes, when compared to teams of the same size from the previous year. 

\subsection*{Case Studies of Multidisciplinarity in COVID-19 Research}
%The field network structure used to calculate author multidisciplinarity reports the relationships between fields of study, with respect to the authors who publish in them. These relationships are altered in COVID-related research (see Figure \ref{fig:interdisciplinary_publication}). To explore the changes to these relationships visually, we make two modifications to the network structure in order to combine all instances of the network (2016-2020) into a single graph: all research published from 2016 to 2019 is consolidated as pre-COVID research, and all edges are now directed. In our combined graph, a directed edge $(SD_A,SD_B)$ represents an author who published in subdiscipline $SD_A$ pre-COVID-19 and in subdiscipline $SD_B$ during COVID-19. We exclude non-COVID-19 work published in 2020 from the network.

The field of study network structure used to calculate author multidisciplinarity encodes relationships between fields of study, with respect to the authors who publish in them. Since these relationships are altered in COVID-related research, we propose a modified network structure to explore the changes to these relationships visually, and to highlight interesting case studies of multidisciplinary research in the COVID-19 literature. In this modified network structure, COVID-related research articles contribute directed edges $(SD_A,SD_B)$ to the graph, for all sub-disciplines $SD_A$ in which the authors publish in their pre-2020 work, and all sub-disciplines $SD_B$ which relate to the article. For example, an edge between the pair of sub-disciplines 'Machine Learning' and 'Radiology' represents an author who published in the field of 'Machine Learning' in their pre-2020 work (2016--2019), publishing COVID-19 research in the field of 'Radiology'. We produce networks of this structure from different subsets of COVID-related research articles, which we will visualise using flow diagrams, where the pre-2020 sub-disciplines are on the left and the COVID-related disciplines are on the right.

% Figures \ref{fig:chord_virology}, \ref{fig:chord_computer_science}, \ref{fig:chord_materials_science}, and \ref{fig:chord_development_economics} are four examples of such networks, produced from COVID-related research in the fields of Virology, Computer Science, Materials Science, and Development Economics respectively. To provide a legible visualisation of the strongest trends, each FoS network shows only the 50 edges with the greatest weights. We choose Virology as a case study as the largest subset in COVID-related research, while Computer Science and Materials Science show considerable increases in author multidisciplinarity in 2020 (see figure \ref{fig:interdisciplinary_publication}), and Development Economics is found to have a very diverse set of contributing disciplines. 

%BS. Once again I've commented out the "results" stuff because I think it needs to be shifted to the Results section. It is conceivable that I am being too anal about this but I'm only coming to fully appreciate and understand the very regimented structure of research articles in at least some disciplines and I've been caught by this myself recently. If you feel strongly that I am wrong or that my suggestion damages the paper then happy to discuss.

\section*{Results}

%Up to three levels of \textbf{subheading} are permitted. Subheadings should not be numbered.

\subsection*{Research Team Size and Collaboration Index}
\begin{figure}[ht]
	\centering
	\includegraphics[width=0.9\linewidth]{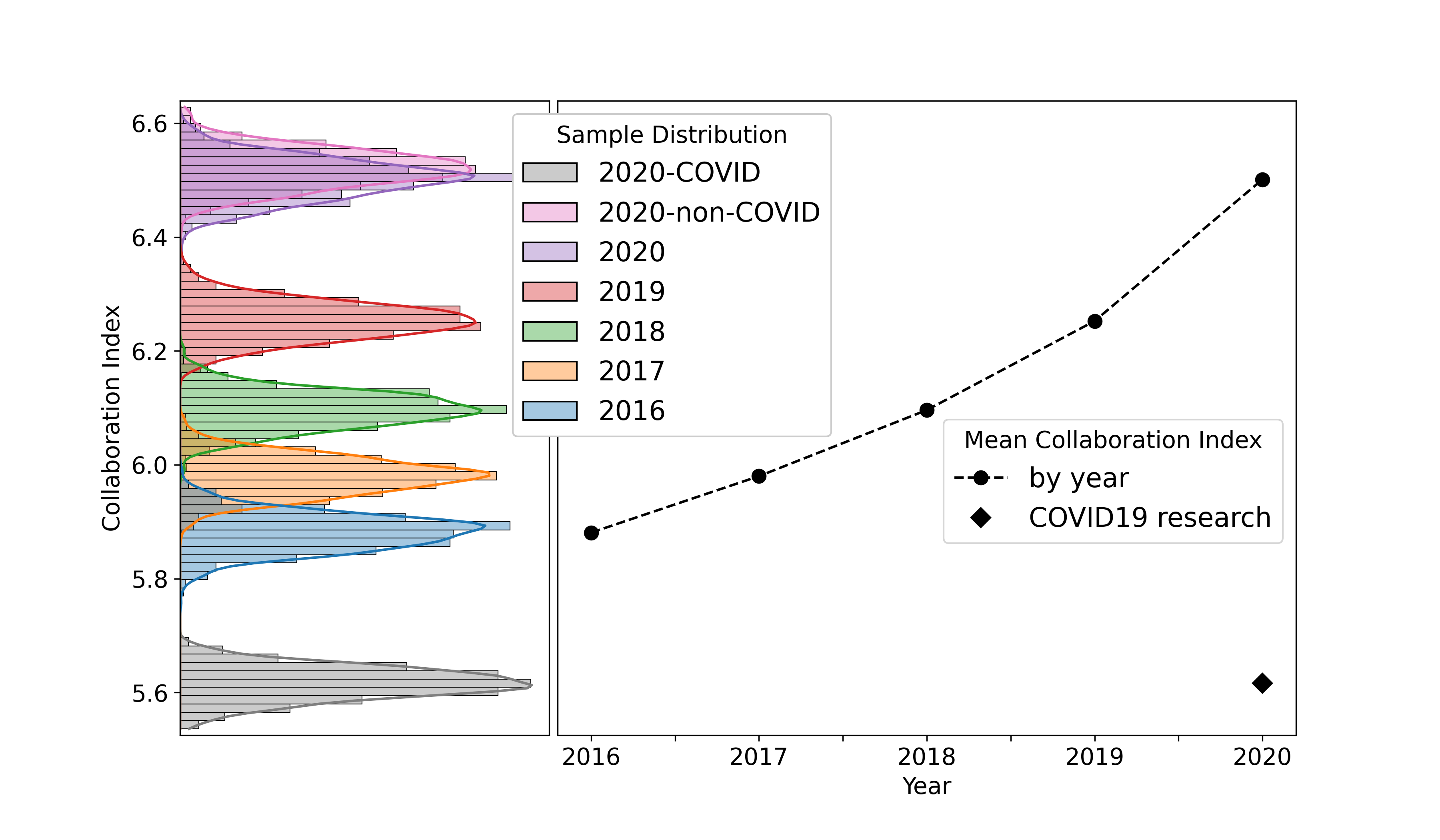}
	\caption{Collaboration Index distributions for samples of 50,000 research articles from different portions of the dataset. 1,000 samples are taken from each year (2016-2020). Collaboration index increases annually, $r^2 = 0.94$, and the CI for COVID-19 articles is significantly less that the CI associated with non-COVID 2020 research; in fact the mean COVID-19 CI is 25 standard deviations below the mean of of non-COVID samples taken from 2020. Thus, research teams publishing COVID-19 research are significantly smaller than expected for research teams in 2020 containing the same authors.}
	\label{fig:collaboration_index}
\end{figure}

Figure \ref{fig:collaboration_index} reports the mean Collaboration Index for the samples of 50,000 research papers taken from each year in the dataset. Mean values for samples of COVID-19 research articles are also included. The Collaboration Index increases year-on-year, indicating a move towards larger research teams. This trend has been noted across many disciplines of academic research \cite{porter_2008,lariviere_2014,leahey_2016}. 

COVID-19 research presents with a very different CI (approximately 5.6), however, indicating that COVID-19 research teams are significantly smaller than expected for research conducted by the same authors in 2020. This result is robust with respect to re-sampling size and in the supplementary materials that accompany this article (see Supplementary Figure 1) we report comparable results using sample sizes $n=$ 10,000 and $n=$ 100,000.

\subsection*{Author Multidisciplinary Publication}
\begin{figure}[!t]
	\centering
	\includegraphics[width=\linewidth]{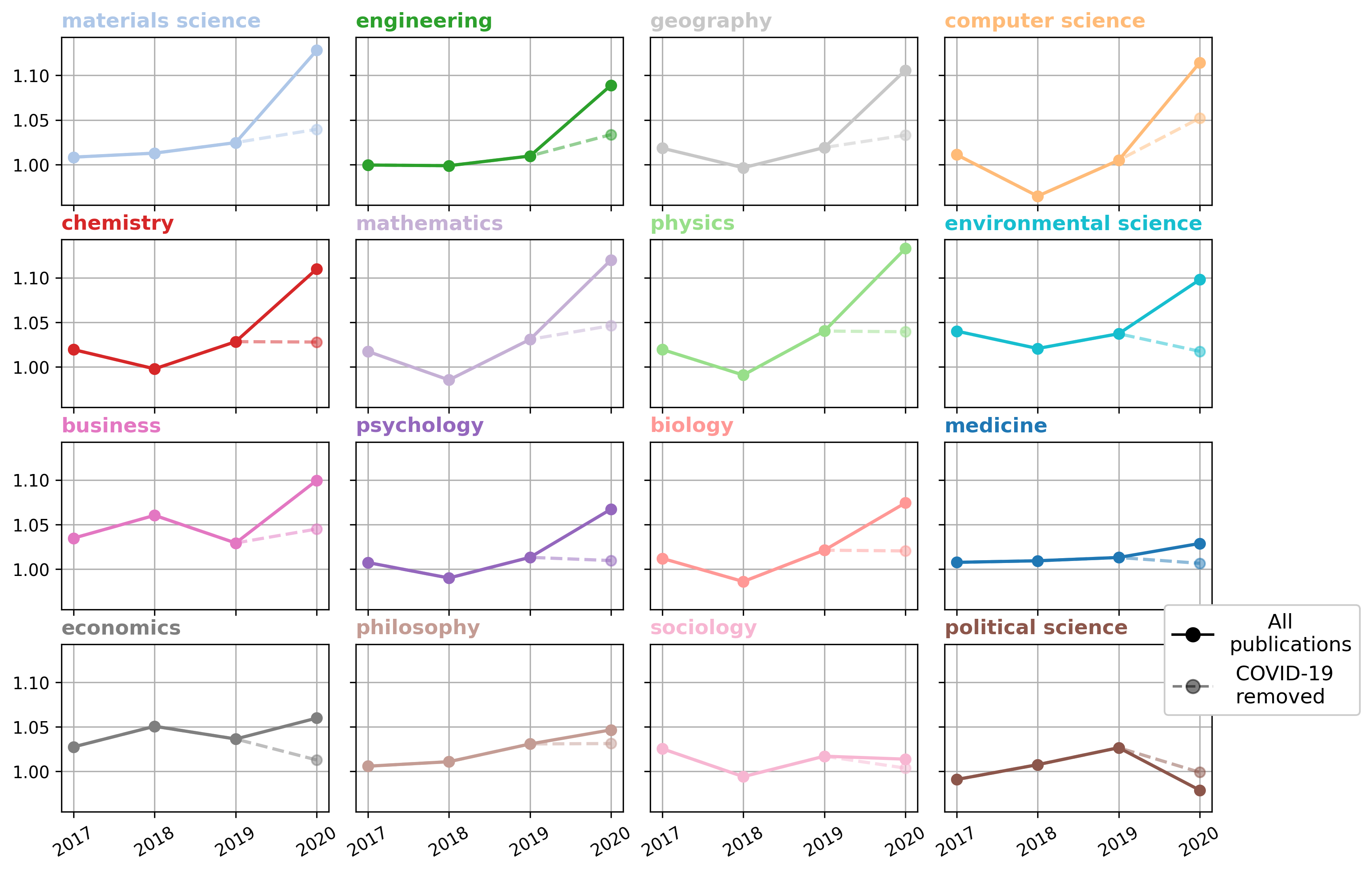}
	\caption{Odds ratio effect sizes for the proportion of links \textit{between} disciplines when compared with the previous year. A score of 1 indicates that authors are no more likely to publish in other disciplines than they were in the previous year.}
	\label{fig:interdisciplinary_publication}
\end{figure}

We quantify author multidisciplinarity in a year of research by measuring the proportion of the total number of edges in an author-FoS network that are \textit{between} communities (i.e., disciplines). We find that this proportion is increasing slowly over time when we produce FoS networks for each year in our data ($slope = 0.2\%, r^2 = 0.98$).
Figure \ref{fig:interdisciplinary_publication} reports the odds ratio effect size when the proportion of the edges that are \textit{between} communities in a given year is compared with that of the previous year. These scores are reported for each community. In the case of 2020 we also report the odds ratio achieved when we compare 2019 with 2020-non-COVID research i.e., after we remove COVID-19 research from the graph. Figure \ref{fig:interdisciplinary_publication} shows an increase in multidisciplinary publication in 2020 across almost all disciplines. The increase in author multidisciplinarity is much greater when we include COVID-19 research in the graph. Despite representing less than 20\% of the work published in 2020, COVID-19 research contributes greatly to the proportion inter-disciplinary edges in the FoS network. 

\subsection*{Research Team Disciplinary Diversity}
\begin{figure}[!t]
	\centering
	\includegraphics[width=0.8\linewidth]{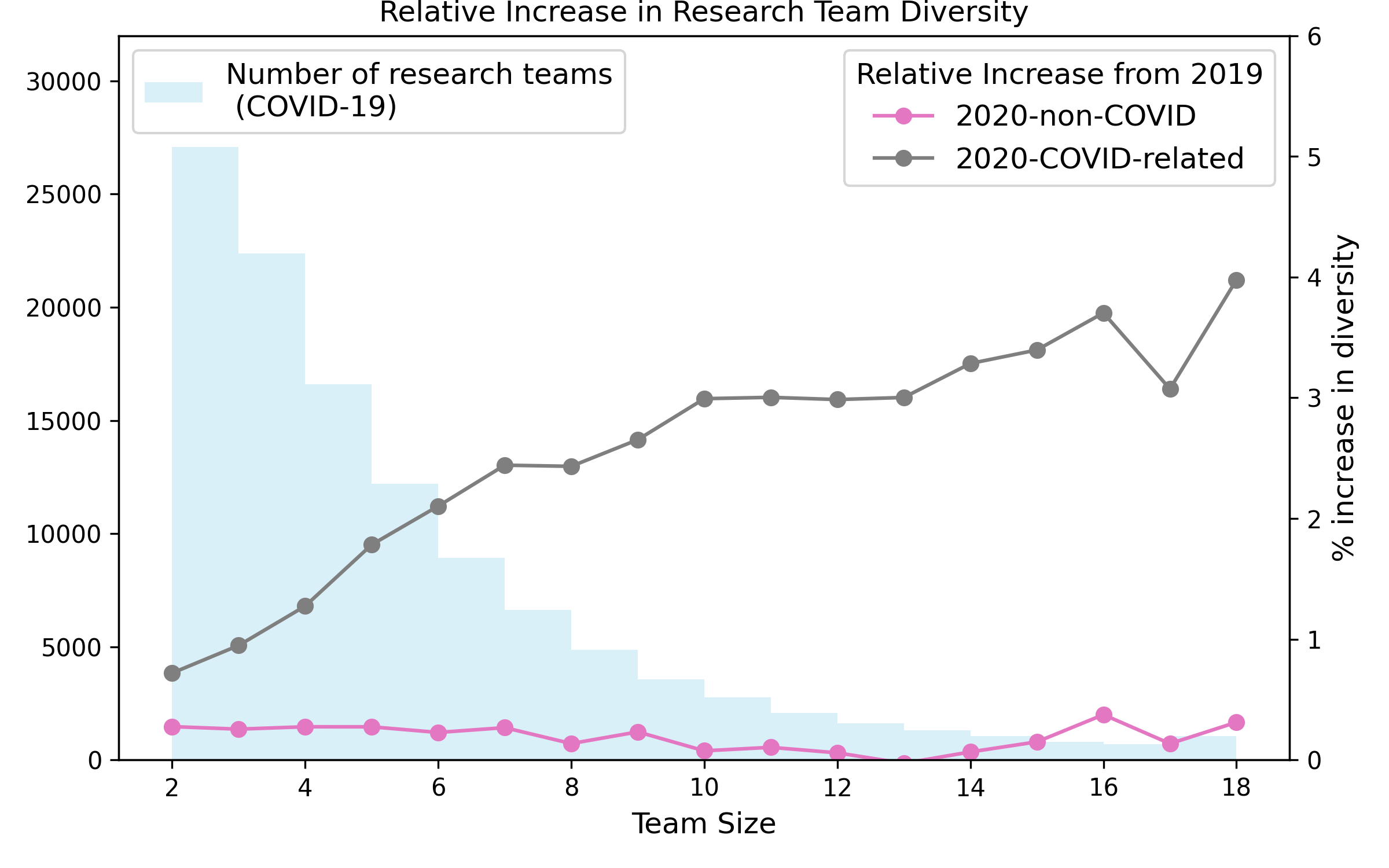}
	\caption{Percentage increase in mean research team diversity for research published in 2020, for teams of increasing size, compared to research published by the same authors in 2019. The distribution of research team sizes is also shown.}
	% independent t-test 
	\label{fig:research_team_diversity}
\end{figure}

When we compare authors by their publication backgrounds, encoded as publication vectors, we find COVID-19 research teams to be more diverse than equivalently-sized research teams who published before 2020. Figure \ref{fig:research_team_diversity} presents the relative increase in mean research team disciplinary diversity for different team sizes, when research teams from 2020 are compared with teams from 2019. We divide 2020 research into two sets: (i) 2020-COVID-related; (ii) 2020-non-COVID research and report relative increases in team diversity for each set. Independent \emph{t} tests show COVID-19 research teams to be significantly more diverse than both pre-2020 and 2020-non-COVID research teams of the same size ($p < 0.01$, see Supplemental Table 4). 

\section*{Discussion}

%The Discussion should be succinct and must not contain subheadings.
%TODO (BS): I think we need to better explain how to interpret the Chord diagrams. For instance, I dont think it will be obvious to the reader that "Figure \ref{fig:chord_virology} shows the intersection between Medicine, Biology and Chemistry in COVID-19 research relating to Virology. Sub-disciplines Molecular Biology, Biochemistry, Immunology, Pathology and Virology all appear closely related in this graph."

Despite the recent trend towards larger, more collaborative research teams \cite{porter_2008,lariviere_2014,leahey_2016,feng_2020}, COVID-19 research appears to have significantly fewer authors than other publications by the same researchers, during 2020. This may be a concerning finding amid evidence that larger teams produce more impactful scientific research \cite{lariviere_2014}: it may have limited the value of the research produced, notwithstanding the incredible achievements that have been made, or it may be a reality of working under the constraints of a global pandemic. % additional references here.

We do see some examples of larger teams and their greater potential for research impact in our analysis: 20\% of COVID-19 research papers have more than 8 listed authors and this portion of the dataset accounts for over 60 of the 100 most cited publications relating to the coronavirus. Yet, the majority of COVID-19 research papers (53\%) have 4 or fewer authors. We find no evidence that the reduced Collaboration Index of COVID-19 research is due to working conditions and restrictions during the pandemic. Despite a global shift towards remote work, research in 2020 continues the recent trend of increasing collaboration. The preference for smaller research teams appears to be specific to COVID-19 research and not simply a factor of research during COVID-19.  

The prevalence of smaller research teams is important to understand about COVID-19 research. Smaller teams have been shown to play a different role to larger teams in both research and technology \cite{wu_2019}.
In an analysis of research collaborations, Wu et al. show that small research teams can disrupt science and technology by exploring and amplifying promising ideas from older and less-popular work, while large teams develop on recent successes by solving acknowledged problems \cite{wu_2019}. The definition by Wu et al. of disruptive articles relates closely to the metric of betweenness centrality for citation networks. 
That is, \textit{disruptive} papers can connect otherwise separate communities in a research network. We find some evidence that COVID-19 research may be increasing the connectivity between disciplines, as authors are more likely to publish across multiple fields and research teams are more diverse. A trend towards greater levels of multidisciplinary collaboration has been identified in many scientific disciplines \cite{porter_2008}. 
This trend is evident in the non-COVID-19 portions of our dataset. Research teams of fewer than 10 members publishing in 2020 exhibit greater disciplinary diversity than similarly-sized teams publishing in 2019, for example. Likewise, the number of authors publishing in multiple disciplines is increasing steadily year-on-year. In COVID-19 research, the increase in multidisciplinarity (of both teams and individuals) exceeds the established trend. This may be evidence of the disruptive nature of COVID-related research. Below, we use flow diagrams to explore author multidisciplinarity in specific topics in the COVID-related research dataset. 

\begin{figure}[ht]
	\centering
	\includegraphics[width=\linewidth]{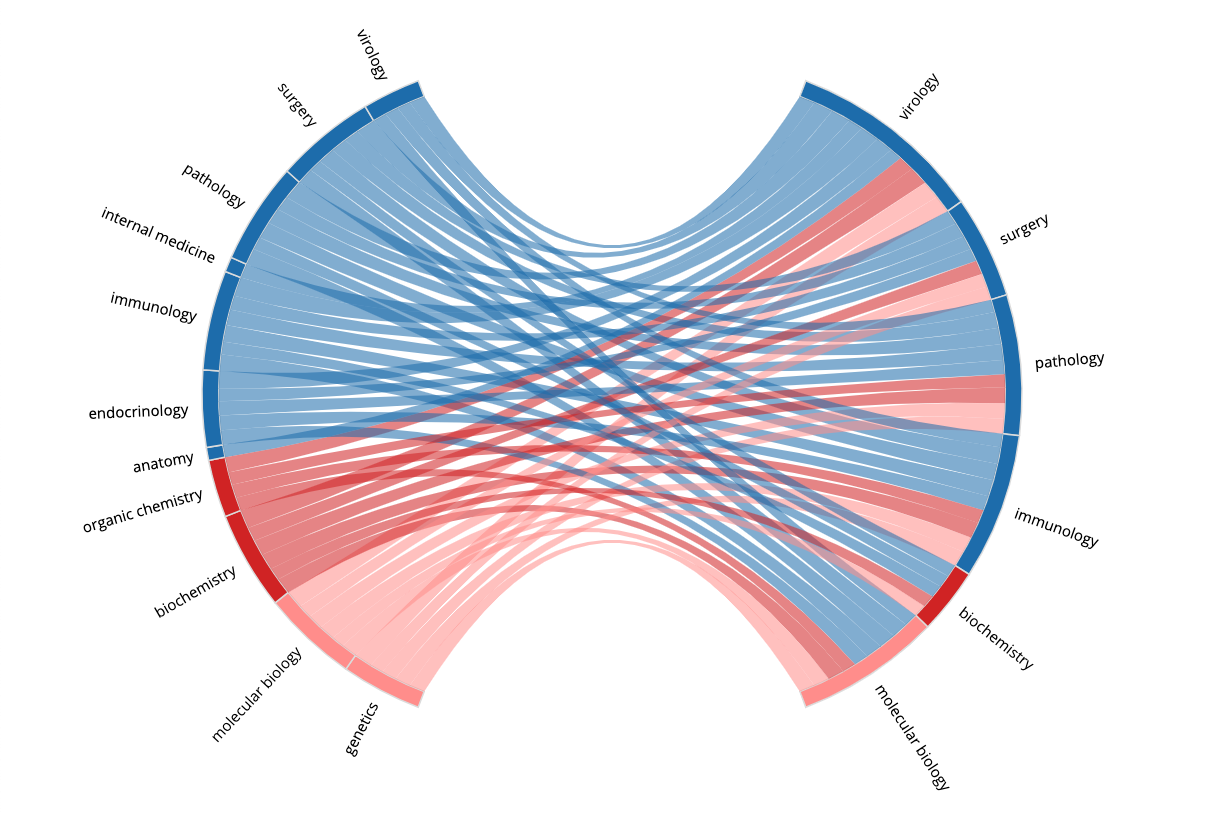}
	\caption{Author Multidisciplinarity in Virology Research in COVID-19. The graph relates an author's research background to the fields they publish in COVID-related articles. This network is produced from 22,561 COVID-related research papers which were assigned the MAG field `Virology'. Pre-COVID sub-disciplines (common in the research backgrounds of the authors) are shown on the left and COVID-related sub-disciplines (common in the article subset) are shown on the right. Sub-disciplines are coloured by their parent disciplines and edges are assigned the colour of the pre-2020 node.
	Edges are weighted by the numbers of authors who published in both of the corresponding sub-disciplines. The bi-gram terms which occurred most frequently in the titles of these papers were: \textit{COVID-19 pandemic, coronavirus disease, SARS-CoV-2 infection} and \textit{novel coronavirus}. (See Supplementary Table 5).}
	\label{fig:chord_virology}
\end{figure}

\begin{figure}[ht]
	\centering
	\includegraphics[width=\linewidth]{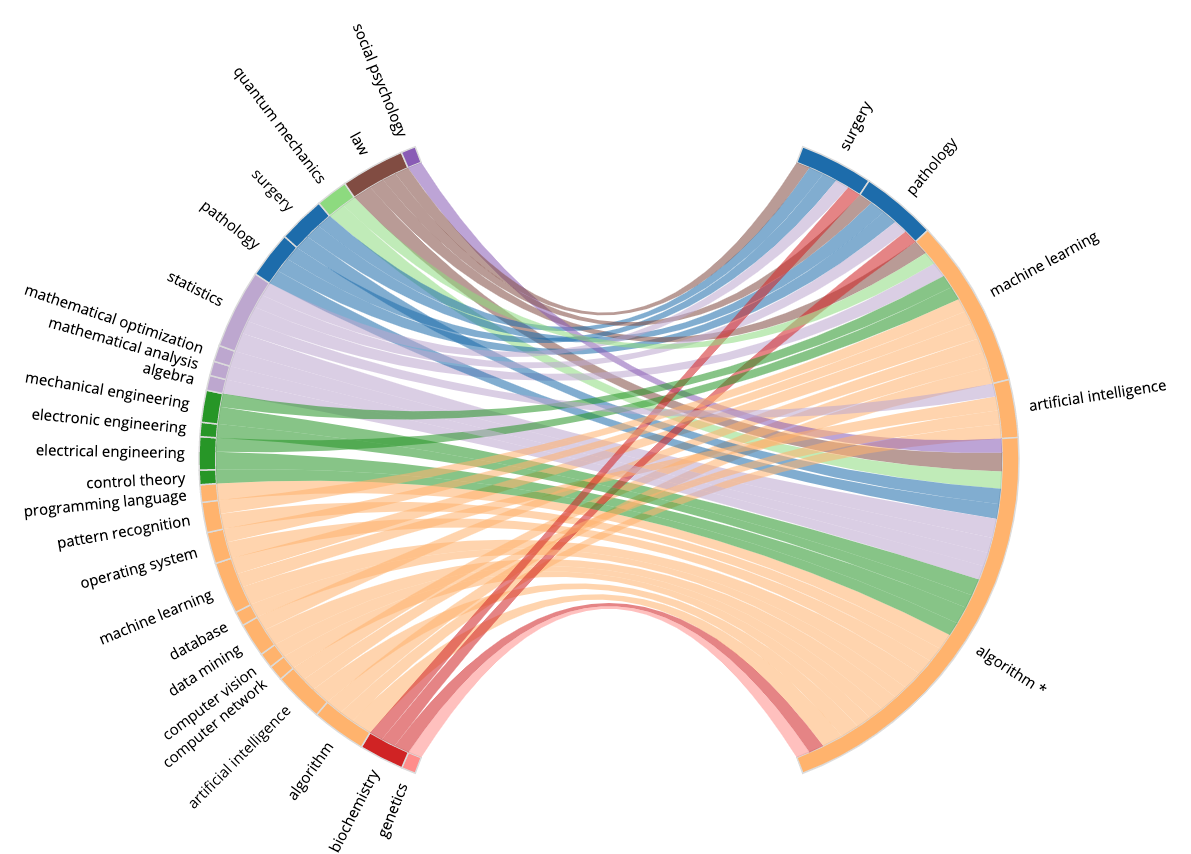}
	\caption{Author Multidisciplinarity in Computer Science Research in COVID-19. This network is produced from 9,004 COVID-related research papers which were attributed the MAG field `Computer Science'. The bi-gram terms which occurred most frequently in the titles of these papers were: \textit{COVID-19 pandemic, deep learning, neural network, machine learning, contact tracing} and \textit{chest x-ray}. *The MAG sub-discipline `Algorithm' is a level 1 parent for any algorithms identified in the fields of study. The most frequently occurring children of the Algorithm field in this subset are \textit{`artificial neural network', `cluster analysis', `inference',} and \textit{`support vector machine'}. (See Supplementary Table 6).}
	\label{fig:chord_computer_science}
\end{figure}

\begin{figure}[ht]
	\centering
	\includegraphics[width=\linewidth]{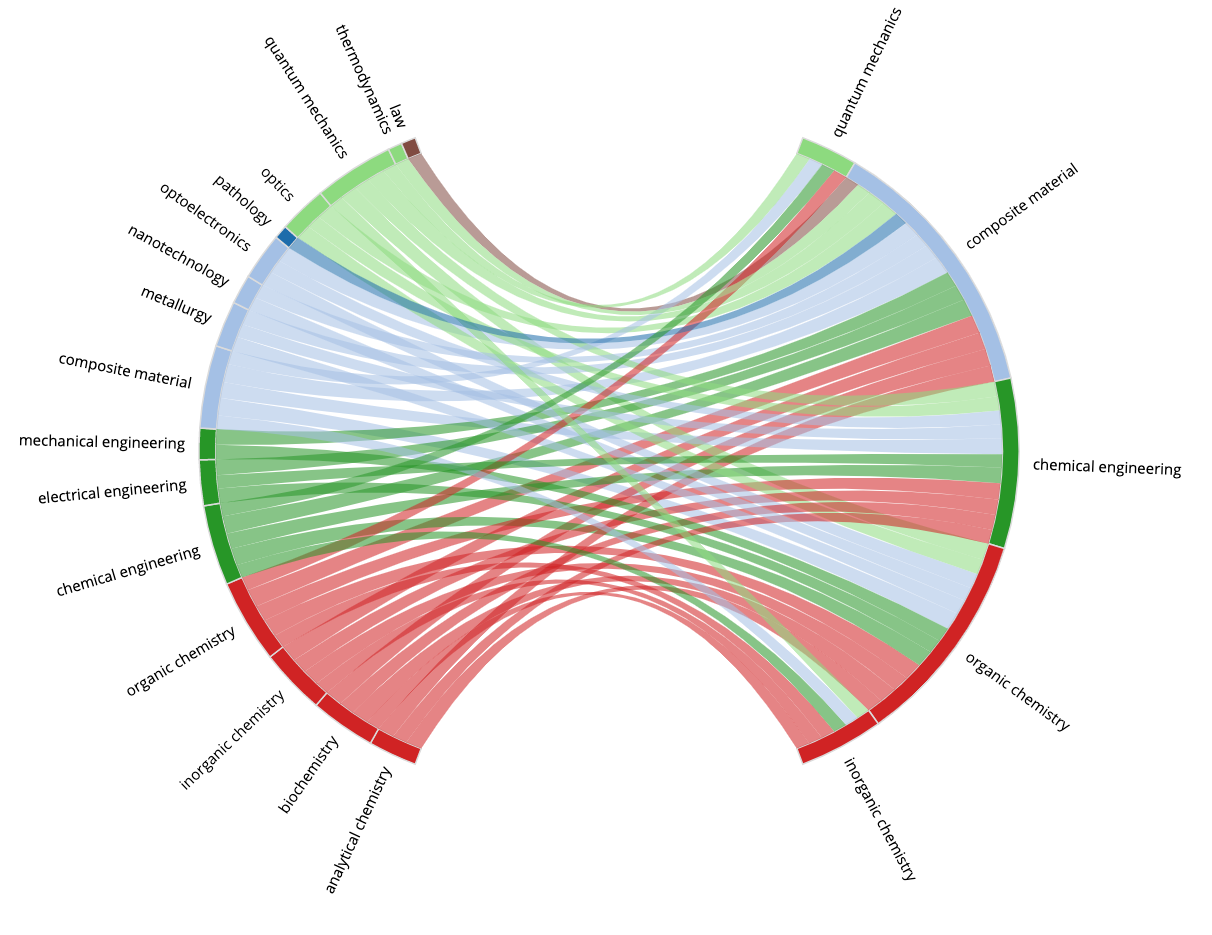}
	\caption{Author Multidisciplinarity in Materials Science Research in COVID-19. This network is produced from 1,229 COVID-related research papers which were attributed the MAG field `Materials Science'. The bi-gram terms which occurred most frequently in the titles of these papers were: \textit{filtration efficiency, additive manufacturing}, and \textit{face mask}. (See Supplementary Table 7).}
	\label{fig:chord_materials_science}
\end{figure}

\begin{figure}[ht]
	\centering
	\includegraphics[width=\linewidth]{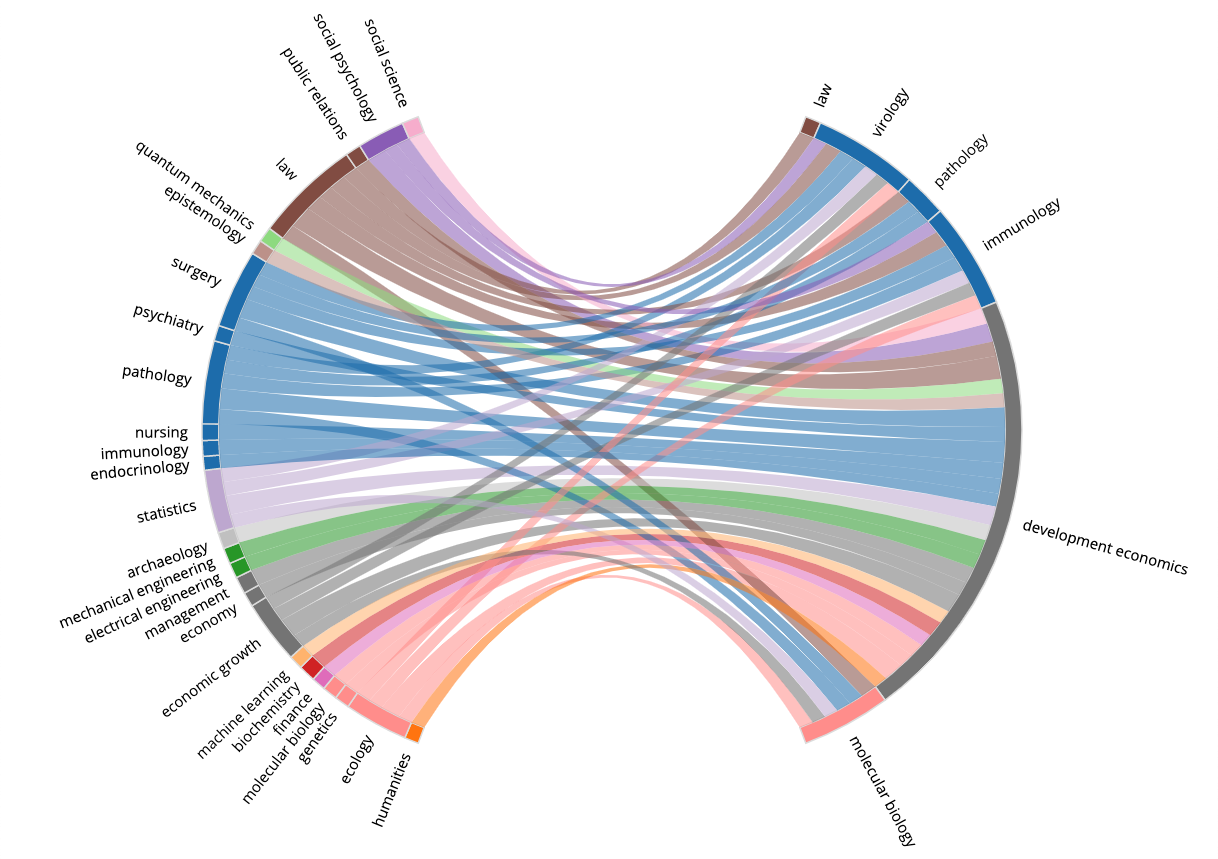}
	\caption{Author Multidisciplinarity in Development Economics Research in COVID-19. The graph relates an author's research background to the fields they publish in COVID-related articles. This network is produced from 1,564 COVID-related research papers which were attributed the MAG field `Development Economics'. The most cited articles in this subset concern studies of the socio-economic implications and effects of the pandemic globally \cite{nicola_2020,walker_2020}, and of health inequity in low- and middle-income countries \cite{patel_2020,wang_2020_equity}. (See Supplementary Table 8).}
	\label{fig:chord_development_economics}
\end{figure}

Figures \ref{fig:chord_virology}, \ref{fig:chord_computer_science}, \ref{fig:chord_materials_science} and \ref{fig:chord_development_economics} present four selected case studies of author multidisciplinarity in COVID-related research in 2020. To provide a clear visualisation of the strongest trends that exist, each FoS network shows only the 50 edges with the greatest weights. We choose Virology as a case study because it is largest subset in COVID-related research, while Computer Science and Materials Science were chosen to show considerable increases in author multidisciplinarity in 2020 (see figure \ref{fig:interdisciplinary_publication}), and Development Economics presents with a very diverse set of contributing disciplines. For example, Figure \ref{fig:chord_virology} shows the intersection between Medicine, Biology and Chemistry in COVID-19 research relating to Virology. Sub-disciplines Molecular Biology, Biochemistry, Immunology, and Virology all appear closely related in this graph. They are strongly interconnected, indicating many instances of authors publishing between disciplines and each acts as both a source and as a destination in the network, as authors who publish in any of these sub-disciplines prior to COVID-19 are likely to publish in the others during COVID-19. % virology/pathology exists at the intersection between these fields - this is not specific to COVID-19 
Figure \ref{fig:chord_computer_science} illustrates the multidisciplinary nature of Computer Science research in COVID-19. Unlike the Virology graph in figure \ref{fig:chord_virology}, there are only two destinations in this network: Computer Science and Medicine. Computer Science research in the COVID-19 dataset is primarily focused on Machine Learning solutions to automating COVID-19 detection from medical images \cite{nguyen_2021} (See Supplementary Table 6(a)). This effort is evident in the graph, as Computer Science research in COVID-19 is most commonly characterised within the sub-disciplines Machine Learning, Artificial Intelligence, Pathology, Surgery and Algorithm. 
Also evident is the multidisciplinary nature of the effort, as researchers with backgrounds in many of the S.T.E.M. fields are shown to contribute. 
Figure \ref{fig:chord_materials_science} reports the FoS network for COVID-19 research relating to Materials Science. The graphs illustrates an intersection between the fields of Physics, Chemistry, Engineering and Materials Science as researchers from each of these disciplines contributes to coronavirus research. Many of the most cited articles in this subset relate to airborne particles and the efficacy of face masks \cite{lustig_2020}, along with the use of electrochemical biosensors for pathogen detection \cite{cesewski_2020} (See Supplementary Table 7(a)).
Figure \ref{fig:chord_development_economics} presents the FoS network for the COVID-19 related research papers in the field of Development Economics. Some of the most cited articles in this subset concern studies of the socio-economic implications and effects of the pandemic globally \cite{nicola_2020,walker_2020}, and of health inequity in low- and middle-income countries \cite{patel_2020, wang_2020_equity} (See Supplementary Table 8(a)). Research in this subset is characterised by the diverse set of sub-disciplines shown on the left of the figure, as authors with backgrounds in social science, social psychology, medicine, statistics, economics, and biology are all found to contribute.

The methods outlined in this work could be applied in future scientometric analyses to assess and visualise multidisciplinarity in a body of research. This may be of interest to researchers seeking to understand the evolution of their own field of study, or to funding agencies who recognise the established benefits of multidisciplinary collaboration. In the case of this work, we show COVID-19 research teams to be smaller yet more multidisciplinary than non-COVID-19 teams.
It is suggested in early work that authors publishing COVID-19 research favoured smaller, less international collaborations in order to reduce co-ordination costs and contribute to the public health effort sooner \cite{fry_2020}. We would like to elaborate on this characterisation of collaboration in COVID-19 research; adding that authors sought to minimise the limitations of working in smaller teams by collaborating with scientists from diverse research backgrounds. That is to say, in the urgency of the pandemic, scientists favour smaller, more multidisciplinary research teams in order to collaborate more efficiently.

\bibliography{references}

% \noindent LaTeX formats citations and references automatically using the bibliography records in your .bib file, which you can edit via the project menu. Use the cite command for an inline citation, e.g.  \cite{Hao:gidmaps:2014}.

% For data citations of datasets uploaded to e.g. \emph{figshare}, please use the \verb|howpublished| option in the bib entry to specify the platform and the link, as in the \verb|Hao:gidmaps:2014| example in the sample bibliography file.

% \section*{Acknowledgements (not compulsory)}

% Acknowledgements should be brief, and should not include thanks to anonymous referees and editors, or effusive comments. Grant or contribution numbers may be acknowledged.

%\section*{Author contributions statement}

%All authors collaborated to design experiments. E.C. collected data, conducted experiments and wrote the main manuscript. D.G. and B.S. analysed results and edited the manuscript.
% Must include all authors, identified by initials, for example:
% A.A. conceived the experiment(s),  A.A. and B.A. conducted the experiment(s), C.A. and D.A. analysed the results.  All authors reviewed the manuscript. 

\section*{Additional information}
\subsection*{Acknowledgments}
This research was supported by Science Foundation Ireland (SFI) under Grant Number SFI/12/RC/2289\_P2.
%\noindent\textbf{Competing interests:}

\subsection*{Data availability}
The data used in our study can reproduced from the set of Microsoft Academic Graph article IDs, which will be made available upon request.
% To include, in this order: \textbf{Accession codes} (where applicable);  (mandatory statement). 

% The corresponding author is responsible for submitting a \href{http://www.nature.com/srep/policies/index.html#competing}{competing interests statement} on behalf of all authors of the paper. This statement must be included in the submitted article file.

% \begin{figure}[ht]
% \centering
% \includegraphics[width=\linewidth]{stream}
% \caption{Legend (350 words max). Example legend text.}
% \label{fig:stream}
% \end{figure}

% \begin{table}[ht]
% \centering
% \begin{tabular}{|l|l|l|}
% \hline
% Condition & n & p \\
% \hline
% A & 5 & 0.1 \\
% \hline
% B & 10 & 0.01 \\
% \hline
% \end{tabular}
% \caption{\label{tab:example}Legend (350 words max). Example legend text.}
% \end{table}

% Figures and tables can be referenced in LaTeX using the ref command, e.g. Figure \ref{fig:stream} and Table \ref{tab:example}.

\end{document}

\section*{Supplemental Materials}

\begin{table}[h]
	\centering
	\begin{tabular}{lrrr}
        \toprule
        {} & Articles & Articles (explicit) & Authors per paper \\
        \midrule
        medicine              & 149,111 & 93,350 &     6.2 \\
        biology               & 131,408 & 12,454 &     6.2 \\
        chemistry             &  54,808 &  5,357 &     7.1 \\
        economics             &  50,389 &  1,357 &     5.2 \\
        psychology            &  49,871 & 10,184 &     5.6 \\
        computer science      &  47,275 &  9,004 &     5.4 \\
        political science     &  45,146 &  8,739 &     5.0 \\
        engineering           &  38,521 &    295 &     5.6 \\
        sociology             &  32,721 &  2,556 &     5.2 \\
        mathematics           &  27,679 &  1,321 &     5.7 \\
        physics               &  26,898 &    576 &     6.1 \\
        philosophy            &  22,442 &    403 &     4.6 \\
        geography             &  14,922 &  7,071 &     5.1 \\
        business              &  14,718 &  6,983 &     4.2 \\
        history               &  11,972 &  2,225 &     4.7 \\
        art                   &  11,519 &    271 &     4.8 \\
        materials science     &   8,123 &  1,229 &     6.0 \\
        geology               &   6,404 &     95 &     5.4 \\
        environmental science &   5,513 &  1,245 &     5.1 \\
        \bottomrule
    \end{tabular}
	\caption{Summary of \emph{2020-COVID-related research} analysed in this work, in terms of the number of research articles and mean number of authors per paper for each discipline. In the case of each discipline, we include any articles that are assigned to fields of study that are within the discipline, i.e., the discipline is a parent of any of the smaller fields of study to which the paper is assigned. We also report the number of articles that are explicitly assigned to that discipline, i.e., the discipline is included in the original set of fields of study attributed to the paper.}
\end{table}

\begin{table}[h]
	\centering
    \begin{tabular}{lrrr}
        \toprule
        {} & Articles & Articles (explicit) & Authors per paper \\
        \midrule
        medicine              & 915,474 & 581,147 &     8.1 \\
        biology               & 865,190 & 105,707 &     8.3 \\
        chemistry             & 551,393 & 105,925 &     8.8 \\
        engineering           & 358,897 &   2,774 &     7.3 \\
        computer science      & 328,293 & 111,965 &     6.9 \\
        psychology            & 320,474 &  74,131 &     7.0 \\
        physics               & 284,391 &  21,023 &     8.6 \\
        mathematics           & 276,038 &  16,153 &     7.0 \\
        political science     & 229,465 &  18,415 &     6.4 \\
        economics             & 226,455 &   9,727 &     6.2 \\
        materials science     & 173,926 &  62,646 &     7.8 \\
        sociology             & 170,212 &  12,583 &     6.5 \\
        philosophy            & 114,323 &   1,041 &     6.1 \\
        geography             &  99,992 &  15,512 &     7.1 \\
        geology               &  76,014 &   5,154 &     7.8 \\
        art                   &  74,140 &   1,328 &     6.8 \\
        business              &  72,259 &  27,964 &     5.1 \\
        history               &  59,372 &   2,492 &     6.4 \\
        environmental science &  48,295 &  22,300 &     6.8 \\
        \bottomrule
    \end{tabular}
    \caption{Summary of \emph{2020-non-COVID research} analysed in this work, in terms of the number of research articles and mean number of authors per paper for each discipline.}

\end{table}

\begin{table}[h]
	\centering
    \begin{tabular}{lrrr}
        {} & Articles & Articles (explicit) & Authors per paper \\
        \midrule
        medicine              & 2,961,762 & 1,838,989 &     7.3 \\
        biology               & 2,804,222 &   434,859 &     7.5 \\
        chemistry             & 1,833,632 &   323,882 &     8.1 \\
        engineering           & 1,203,777 &    45,147 &     6.7 \\
        psychology            & 1,035,184 &   241,721 &     6.2 \\
        computer science      & 1,033,976 &   336,604 &     6.3 \\
        physics               &   960,508 &    81,679 &     8.2 \\
        mathematics           &   862,481 &    68,010 &     6.3 \\
        political science     &   750,923 &    72,185 &     5.6 \\
        economics             &   745,753 &    43,975 &     5.4 \\
        materials science     &   590,016 &   208,028 &     7.3 \\
        sociology             &   548,850 &    46,088 &     5.6 \\
        philosophy            &   419,432 &    11,681 &     5.4 \\
        geography             &   332,303 &    50,735 &     6.5 \\
        art                   &   280,670 &    12,696 &     5.8 \\
        geology               &   249,208 &    19,768 &     7.3 \\
        business              &   244,328 &    82,362 &     4.5 \\
        history               &   200,924 &     8,940 &     5.8 \\
        environmental science &   157,073 &    64,045 &     6.1 \\
        \bottomrule
    \end{tabular}

    \caption{Summary of \emph{pre-2020 research} analysed in this work, in terms of the number of research articles and mean number of authors per paper for each discipline.}
\end{table}

\newpage 
\begin{figure}
\adjustbox{valign=t}{%
\begin{minipage}[t]{.2\linewidth}
\caption{Sample distributions for Collaboration Index scores for different years and portions of the dataset. The experiment is repeated for different sample sizes. The trend towards greater collaboration is evident regardless of sample size and the COVID-19 research is consistently shown to be significantly less collaborative than other research in 2020.}
\end{minipage}}%
\adjustbox{valign=t}{%
\begin{minipage}[t]{.85\linewidth}
  \begin{subfigure}{\linewidth}
  \centering
  \includegraphics[width=.85\linewidth]{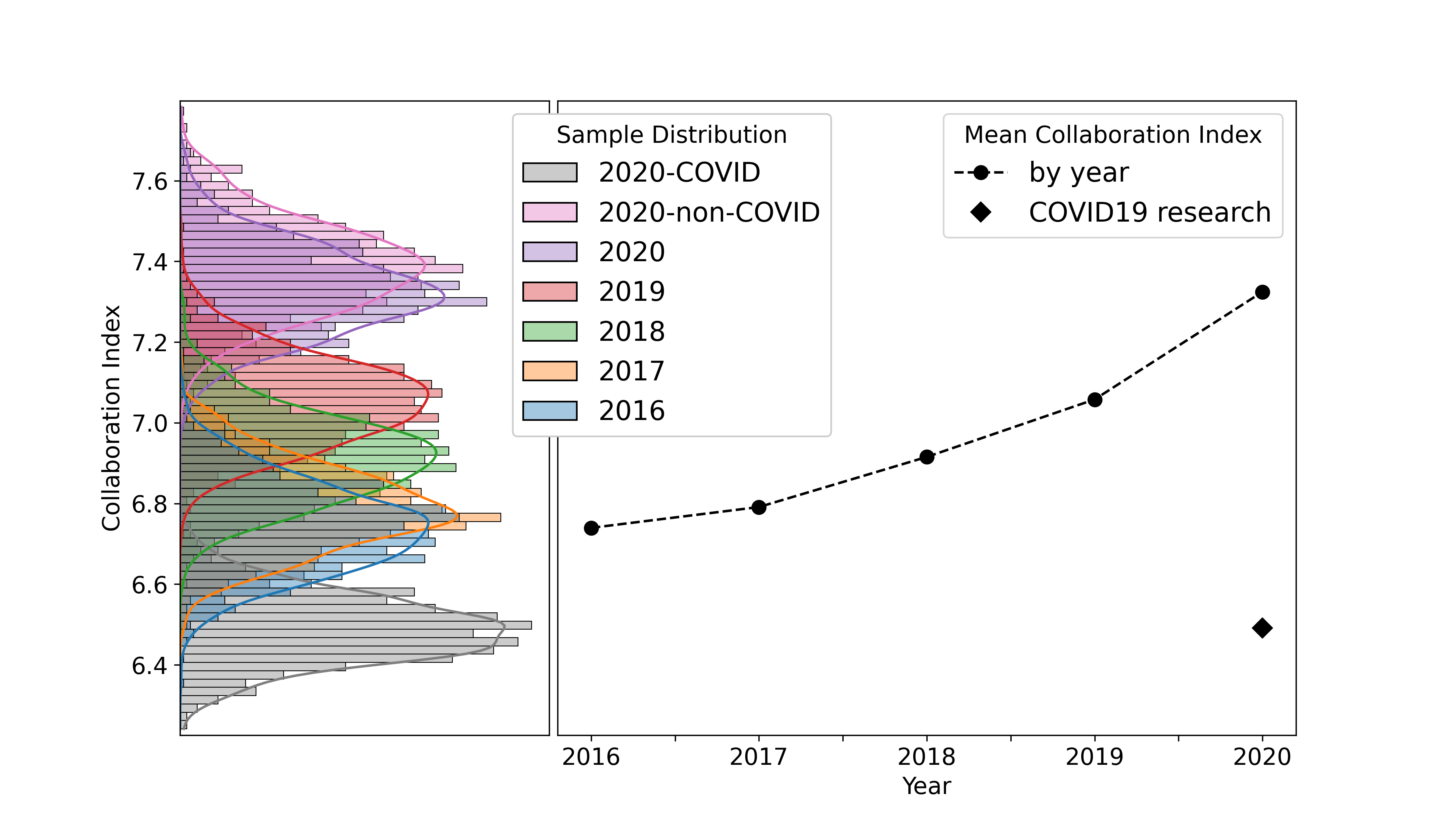}
  \vspace{-1\baselineskip}
  \caption{$n=10,000$}
  \vspace{-2\baselineskip}
  \end{subfigure}\par\bigskip
  \begin{subfigure}{\linewidth}
  \centering
  \includegraphics[width=.85\linewidth]{fig_1_collaboration_index_50.png}
  \vspace{-1\baselineskip}
  \caption{$n=50,000$}
  \vspace{-2\baselineskip}
  \end{subfigure}\par\bigskip
  \begin{subfigure}{\linewidth}
  \centering
  \includegraphics[width=.85\linewidth]{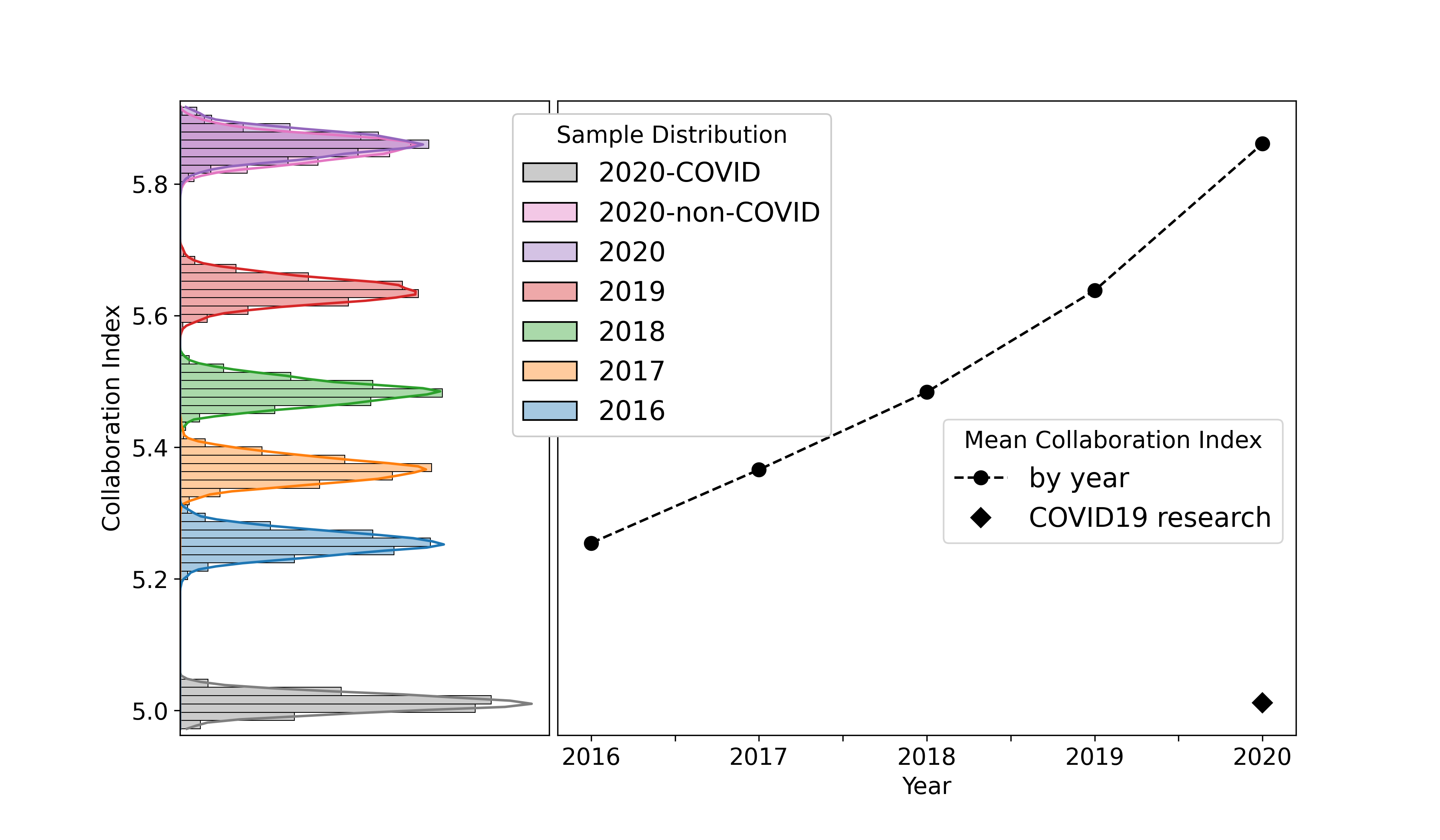}
  \vspace{-1\baselineskip}
  \caption{$n=100,000$}
  \vspace{-1\baselineskip}
  \end{subfigure}
\end{minipage}}
\end{figure}

\begin{table}[h]
	\begin{subtable}[h]{0.48\textwidth}
        \begin{tabular}{lrrrr}
        Team  & {} & {} & mean ± std & mean ± std \\
        size & t &     p &           2020-COVID &  pre-2020 \\
        \midrule
        2         &  6.73 &  8.34e-12* &  0.134 ± 0.15 &  0.128 ± 0.14 \\
        3         & 10.41 &  1.16e-25* &  0.136 ± 0.12 &  0.128 ± 0.11 \\
        4         & 14.14 &  1.20e-45* &  0.135 ± 0.11 &  0.124 ± 0.10 \\
        5         & 19.07 &  2.96e-81* &  0.134 ± 0.10 &  0.118 ± 0.09 \\
        6         & 21.16 &  1.64e-99* &  0.132 ± 0.09 &  0.113 ± 0.08 \\
        7         & 23.06 & 1.30e-117* &  0.130 ± 0.09 &  0.108 ± 0.07 \\
        8         & 20.71 &  2.71e-95* &  0.126 ± 0.08 &  0.104 ± 0.07 \\
        9         & 20.74 &  2.08e-95* &  0.124 ± 0.08 &  0.100 ± 0.07 \\
        10        & 20.99 &  1.51e-97* &  0.125 ± 0.08 &  0.098 ± 0.06 \\
        11        & 18.82 &  7.85e-79* &  0.123 ± 0.08 &  0.096 ± 0.06 \\
        12        & 16.95 &  2.38e-64* &  0.121 ± 0.07 &  0.094 ± 0.06 \\
        13        & 15.71 &  1.61e-55* &  0.120 ± 0.07 &  0.093 ± 0.06 \\
        14        & 15.67 &  3.36e-55* &  0.121 ± 0.07 &  0.091 ± 0.06 \\
        15        & 14.27 &  4.12e-46* &  0.120 ± 0.07 &  0.089 ± 0.06 \\
        16        & 14.95 &  3.16e-50* &  0.121 ± 0.07 &  0.087 ± 0.06 \\
        17        & 11.65 &  2.04e-31* &  0.116 ± 0.07 &  0.088 ± 0.06 \\
        18        & 13.04 &  1.21e-38* &  0.122 ± 0.07 &  0.086 ± 0.06 \\
        19        & 12.40 &  3.93e-35* &  0.119 ± 0.07 &  0.083 ± 0.05 \\
        \bottomrule
        \end{tabular}
		\caption{Alternative hypothesis: 2020-COVID research teams are more diverse than pre-2020.}
	\end{subtable}
	\hspace{\fill}
	\begin{subtable}[h]{0.48\textwidth}
		\begin{tabular}{lrrrr}
        Team & {} & {} & mean ± std & mean ± std \\
        size & t &     p &           2020-COVID &   2020-non-COVID \\
        \midrule
        2         &  4.11 & 1.98e-05* &  0.134 ± 0.15 &  0.131 ± 0.14 \\
        3         &  7.48 & 3.76e-14* &  0.136 ± 0.12 &  0.130 ± 0.11 \\
        4         & 10.86 & 8.93e-28* &  0.135 ± 0.11 &  0.126 ± 0.10 \\
        5         & 15.77 & 2.87e-56* &  0.134 ± 0.10 &  0.121 ± 0.09 \\
        6         & 18.39 & 9.83e-76* &  0.132 ± 0.09 &  0.115 ± 0.08 \\
        7         & 19.88 & 4.93e-88* &  0.130 ± 0.09 &  0.110 ± 0.07 \\
        8         & 19.15 & 9.35e-82* &  0.126 ± 0.08 &  0.106 ± 0.07 \\
        9         & 18.13 & 1.65e-73* &  0.124 ± 0.08 &  0.102 ± 0.07 \\
        10        & 20.15 & 4.70e-90* &  0.125 ± 0.08 &  0.099 ± 0.06 \\
        11        & 17.81 & 8.98e-71* &  0.123 ± 0.08 &  0.097 ± 0.06 \\
        12        & 16.31 & 1.18e-59* &  0.121 ± 0.07 &  0.094 ± 0.06 \\
        13        & 15.74 & 1.16e-55* &  0.120 ± 0.07 &  0.092 ± 0.06 \\
        14        & 14.93 & 3.29e-50* &  0.121 ± 0.07 &  0.092 ± 0.06 \\
        15        & 13.55 & 9.72e-42* &  0.120 ± 0.07 &  0.091 ± 0.06 \\
        16        & 12.53 & 6.14e-36* &  0.121 ± 0.07 &  0.090 ± 0.06 \\
        17        & 10.58 & 3.33e-26* &  0.116 ± 0.07 &  0.089 ± 0.06 \\
        18        & 11.16 & 7.31e-29* &  0.122 ± 0.07 &  0.088 ± 0.06 \\
        19        & 10.30 & 7.60e-25* &  0.119 ± 0.07 &  0.087 ± 0.06 \\
        \bottomrule
        \end{tabular}
		\caption{Alternative hypothesis: 2020-COVID research teams are more diverse than 2020-non-COVID.}
	\end{subtable}
    \newline
    \vspace*{1 cm}
    \newline
	\begin{subtable}[h]{0.48\textwidth}
	    \flushright
		\begin{tabular}{lrrrr}
        Team & {} & {} & mean ± std & mean ± std \\
        size & t &     p &           2020-non-COVID &   pre-2020 \\
        \midrule
        2         &  4.82 & 7.30e-07* &  0.131 ± 0.14 &  0.128 ± 0.14 \\
        3         &  5.72 & 5.22e-09* &  0.130 ± 0.11 &  0.128 ± 0.11 \\
        4         &  6.73 & 8.24e-12* &  0.126 ± 0.10 &  0.124 ± 0.10 \\
        5         &  6.79 & 5.64e-12* &  0.121 ± 0.09 &  0.118 ± 0.09 \\
        6         &  5.54 & 1.50e-08* &  0.115 ± 0.08 &  0.113 ± 0.08 \\
        7         &  6.19 & 3.00e-10* &  0.110 ± 0.07 &  0.108 ± 0.07 \\
        8         &  2.91 & 1.80e-03* &  0.106 ± 0.07 &  0.104 ± 0.07 \\
        9         &  4.59 & 2.27e-06* &  0.102 ± 0.07 &  0.100 ± 0.07 \\
        10        &  1.36 &      0.09 &  0.099 ± 0.06 &  0.098 ± 0.06 \\
        11        &  1.66 &      0.05 &  0.097 ± 0.06 &  0.096 ± 0.06 \\
        12        &  0.85 &      0.20 &  0.094 ± 0.06 &  0.094 ± 0.06 \\
        13        & -0.36 &      0.64 &  0.092 ± 0.06 &  0.093 ± 0.06 \\
        14        &  0.78 &      0.22 &  0.092 ± 0.06 &  0.091 ± 0.06 \\
        15        &  1.58 &      0.06 &  0.091 ± 0.06 &  0.089 ± 0.06 \\
        16        &  3.52 & 2.19e-04* &  0.090 ± 0.06 &  0.087 ± 0.06 \\
        17        &  1.15 &      0.12 &  0.089 ± 0.06 &  0.088 ± 0.06 \\
        18        &  2.45 & 7.09e-03* &  0.088 ± 0.06 &  0.086 ± 0.06 \\
        19        &  3.26 & 5.67e-04* &  0.087 ± 0.06 &  0.083 ± 0.05 \\
        \bottomrule
        \end{tabular}
		\caption{Alternative hypothesis: 2020-non-COVID research teams are more diverse than pre-2020.}
	\end{subtable}
	\caption{Independent \emph{t} tests comparing research team diversities. We compare diversity scores for teams of equal sizes with the null hypothesis that the mean diversity in each distribution is the same. Each test is one-tailed with the alternative hypothesis stated in the sub-caption. Values significant at $p < 0.01$ are marked with a *. COVID research teams of all sizes are shown to be significantly more diverse than non-COVID baselines (see tables 4(a) and 4(b)).}
\end{table}

\begin{table}[h]
	\centering
	\small
	\setlength\tabcolsep{2pt}
	\begin{subtable}[h]{\textwidth}
		\centering
		\begin{tabular}{llr}
			\toprule
			{} & Title                                                                                                       & Citations \\
			\midrule
			1  & A Novel Coronavirus from Patients with Pneumonia in China, 2019.                                            & 6700      \\
			2  & A pneumonia outbreak associated with a new coronavirus of probable bat origin                               & 5562      \\
			3  & Early Transmission Dynamics in Wuhan, China, of Novel Coronavirus-Infected Pneumonia.                       & 5002      \\
			4  & SARS-CoV-2 Cell Entry Depends on ACE2 and TMPRSS2 and Is Blocked by a Clinically Proven Protease Inhibitor. & 4345      \\
			5  & A new coronavirus associated with human respiratory disease in China.                                       & 2783      \\
			6  & Aerosol and Surface Stability of SARS-CoV-2 as Compared with SARS-CoV-1.                                    & 2499      \\
			7  & Cryo-EM structure of the 2019-nCoV spike in the prefusion conformation.                                     & 2321      \\
			8  & Remdesivir and chloroquine effectively inhibit the recently emerged novel coronavirus (2019-nCoV) in vitro. & 2239      \\
			\bottomrule
		\end{tabular}
		\caption{Most cited papers}
	\end{subtable}
	\begin{subtable}[h]{\textwidth}
		\centering
		\begin{tabular}{lllll}
			\toprule
			{} & Title unigrams     & Title bigrams              & Abstract           & Field of Study                                          \\
			\midrule
			1  & covid-19 (11129)   & covid-19 pandemic (1543)   & virus (4822)       & virology (22561)                                        \\
			2  & sars-cov-2 (5321)  & coronavirus disease (1275) & 2 (4595)           & coronavirus disease 2019 (18238)                        \\
			3  & coronavirus (3888) & disease 2019 (1058)        & sars (4168)        & severe acute respiratory syndrome coronavirus 2 (17579) \\
			4  & pandemic (2680)    & sars-cov-2 infection (748) & cov (4135)         & 2019 20 coronavirus outbreak (16990)                    \\
			5  & disease (2058)     & novel coronavirus (647)    & coronavirus (3875) & medicine (15247)                                        \\
			6  & infection (2057)   & covid-19 patient (469)     & 19 (3873)          & pandemic (8790)                                         \\
			7  & patient (1893)     & 2019 covid-19 (410)        & covid (3801)       & betacoronavirus (5784)                                  \\
			8  & virus (1521)       & acute respiratory (384)    & infection (3670)   & coronavirus (5159)                                      \\
			9  & 2019 (1376)        & respiratory syndrome (365) & disease (3564)     & biology (4582)                                          \\
			10 & de (1042)          & severe acute (354)         & respiratory (2746) & coronavirus infections (3210)                           \\
			\bottomrule
		\end{tabular}
		
		\caption{Most frequent terms}
	\end{subtable}
	\caption{\label{tab:virology}Virology papers. A description of 22,561 Virology research articles. Table (a) lists the titles of the most cited papers in the subset. Table (b) lists terms that appear most frequently in the titles and abstracts of the papers, along with the most common MAG FoS. The number of articles with the respective terms/FoS are included in brackets.}
\end{table}

\begin{table}[h]
	\centering
	\small
	\setlength\tabcolsep{2pt}
	\begin{subtable}[h]{\textwidth}
		\centering
		\begin{tabular}{llr}
			\toprule
			{} & Title                                                                                                                   & Citations \\
			\midrule
			1  & Detection of 2019 novel coronavirus (2019-nCoV) by real-time RT-PCR.                                                    & 1899      \\
			2  & Quantifying SARS-CoV-2 transmission suggests epidemic control with digital contact tracing.                             & 680       \\
			3  & COVID-Net: a tailored deep convolutional neural network design for detection of COVID-19 cases from chest X-ray images. & 320       \\
			4  & Diagnosing COVID-19: The Disease and Tools for Detection                                                                & 257       \\
			5  & Covid-19: automatic detection from X-ray images utilizing transfer learning with convolutional neural networks.         & 228       \\
			6  & A deep learning algorithm using CT images to screen for Corona Virus Disease (COVID-19)                                 & 227       \\
			7  & Automatic Detection of Coronavirus Disease (COVID-19) Using X-ray Images and Deep Convolutional Neural Networks.        & 204       \\
			8  & An updated estimation of the risk of transmission of the novel coronavirus (2019-nCov)                                  & 201       \\
			\bottomrule
		\end{tabular}
		\caption{Most cited papers}
	\end{subtable}
	\begin{subtable}[h]{\textwidth}
		\centering
		
		\begin{tabular}{lllll}
			\toprule
			{} & Title unigrams  & Title bigrams                 & Abstract        & Field of Study                                         \\
			\midrule
			1  & covid-19 (2792) & covid-19 pandemic (348)       & model (4908)    & computer science (9004)                                \\
			2  & using (1097)    & deep learning (268)           & 19 (3734)       & coronavirus disease 2019 (3453)                        \\
			3  & learning (1005) & machine learning (261)        & covid (3729)    & artificial intelligence (2081)                         \\
			4  & model (959)     & neural network (260)          & data (3632)     & severe acute respiratory syndrome coronavirus 2 (1287) \\
			5  & data (722)      & contact tracing (179)         & paper (2897)    & 2019 20 coronavirus outbreak (1036)                    \\
			6  & network (702)   & chest x-ray (139)             & approach (2863) & pandemic (1025)                                        \\
			7  & system (569)    & x-ray image (127)             & system (2830)   & machine learning (835)                                 \\
			8  & analysis (566)  & case study (120)              & time (2757)     & deep learning (711)                                    \\
			9  & based (537)     & artificial intelligence (109) & network (2251)  & data science (620)                                     \\
			10 & detection (533) & model covid-19 (101)          & pandemic (2247) & pattern recognition (505)                              \\
			\bottomrule
		\end{tabular}
		
		\caption{Most frequent terms}
	\end{subtable}
	\caption{\label{tab:comp_sci}Computer Science papers. A description of 9,004 Computer Science research articles. Table (a) lists the titles of the most cited papers in the subset. Table (b) lists terms that appear most frequently in the titles and abstracts of the papers, along with the most common MAG FoS. The number of articles with the respective terms/FoS are included in brackets.}
\end{table}

\begin{table}[h]
	\centering
	\small
	\setlength\tabcolsep{2pt}
	\begin{subtable}[h]{\textwidth}
		\centering
		\resizebox{\columnwidth}{!}{\begin{tabular}{llr} 
			\toprule
			{} &                                                                                                                                                  Title &  Citations \\
			\midrule
			1 &                                              The airborne lifetime of small speech droplets and their potential importance in SARS-CoV-2 transmission. &        204 \\
			2 &                                                                                                      Electrochemical biosensors for pathogen detection &         41 \\
			3 &               Household Materials Selection for Homemade Cloth Face Coverings and Their Filtration Efficiency Enhancement with Triboelectric Charging. &         28 \\
			4 &  The versatile biomedical applications of bismuth-based nanoparticles and composites: therapeutic, diagnostic, biosensing, and regenerative properties &         26 \\
			5 &                                                                       DeepCOVIDExplainer: Explainable COVID-19 Predictions Based on Chest X-ray Images &         24 \\
			6 &                                            Green synthesis, biomedical and biotechnological applications of carbon and graphene quantum dots. A review &         22 \\
			7 &                            Modeling ambient temperature and relative humidity sensitivity of respiratory droplets and their role in Covid-19 outbreaks &         21 \\
			8 &                                                                 An overview of functional nanoparticles as novel emerging antiviral therapeutic agents &         20 \\
			\bottomrule
			\end{tabular}}
		\caption{Most cited papers}
	\end{subtable}
	\begin{subtable}[h]{\textwidth}
		\centering
		\begin{tabular}{lllll}
			\toprule
			{} & Title unigrams     & Title bigrams               & Abstract          & Field of Study                                       \\
			\midrule
			1  & effect (88)        & filtration efficiency (14)  & material (404)    & materials science (1229)                             \\
			2  & using (86)         & additive manufacturing (13) & high (403)        & nanotechnology (152)                                 \\
			3  & application (86)   & mechanical property (10)    & surface (402)     & chemical engineering (149)                           \\
			4  & covid-19 (83)      & solar cell (9)              & property (321)    & coronavirus disease 2019 (143)                       \\
			5  & property (70)      & face mask (9)               & application (315) & composite material (135)                             \\
			6  & surface (68)       & respiratory droplet (9)     & effect (312)      & optoelectronics (123)                                \\
			7  & material (62)      & recent advance (8)          & 2 (281)           & severe acute respiratory syndrome coronavirus 2 (93) \\
			8  & based (52)         & covid-19 pandemic (8)       & 1 (275)           & nanoparticle (74)                                    \\
			9  & nanoparticles (47) & 3d printed (8)              & time (269)        & biomedical engineering (73)                          \\
			10 & study (44)         & quantum dot (8)             & different (265)   & 2019 20 coronavirus outbreak (65)                    \\
			\bottomrule
		\end{tabular}
		\caption{Most frequent terms}
	\end{subtable}
	\caption{\label{tab:materials_science}Materials Science papers. A description of 1,229 Materials Science research articles. Table (a) lists the titles of the most cited papers in the subset. Table (b) lists terms that appear most frequently in the titles and abstracts of the papers, along with the most common MAG FoS. The number of articles with the respective terms/FoS are included in brackets.}
\end{table}

\begin{table}[h]
	\centering
	\small
	\setlength\tabcolsep{2pt}
	\begin{subtable}[h]{\textwidth}
		\centering
		\begin{tabular}{llr}
			\toprule
			{} & Title                                                                                                       & Citations \\
			\midrule
			1  & The socio-economic implications of the coronavirus pandemic (COVID-19): A review.                           & 505       \\
			2  & COVID-19: towards controlling of a pandemic.                                                                & 379       \\
			3  & Return of the Coronavirus: 2019-nCoV.                                                                       & 337       \\
			4  & Effective containment explains subexponential growth in recent confirmed COVID-19 cases in China.           & 224       \\
			5  & The effect of large-scale anti-contagion policies on the COVID-19 pandemic.                                 & 185       \\
			6  & Countries test tactics in 'war' against COVID-19.                                                           & 134       \\
			7  & The impact of COVID-19 and strategies for mitigation and suppression in low- and middle-income countries.   & 116       \\
			8  & If the world fails to protect the economy, COVID-19 will damage health not just now but also in the future. & 101       \\
			\bottomrule
		\end{tabular}

		\caption{Most cited papers}
	\end{subtable}
	\begin{subtable}[h]{\textwidth}
		\centering
		\begin{tabular}{lllll}
			\toprule
			{} & Title unigrams  & Title bigrams           & Abstract        & Field of Study                                        \\
			\midrule
			1  & covid-19 (1052) & covid-19 pandemic (247) & covid (1158)    & development economics (1564)                          \\
			2  & pandemic (450)  & impact covid-19 (63)    & 19 (1152)       & coronavirus disease 2019 (1072)                       \\
			3  & health (172)    & response covid-19 (41)  & pandemic (1031) & pandemic (997)                                        \\
			4  & crisis (157)    & covid-19 crisis (41)    & country (816)   & 2019 20 coronavirus outbreak (573)                    \\
			5  & global (149)    & social distancing (35)  & health (617)    & severe acute respiratory syndrome coronavirus 2 (531) \\
			6  & impact (148)    & covid-19 outbreak (34)  & disease (535)   & business (507)                                        \\
			7  & response (123)  & spread covid-19 (29)    & economic (533)  & political science (447)                               \\
			8  & economic (119)  & pandemic covid-19 (29)  & impact (511)    & geography (282)                                       \\
			9  & social (111)    & public health (27)      & social (480)    & public health (248)                                   \\
			10 & policy (103)    & covid-19 epidemic (25)  & policy (456)    & outbreak (210)                                        \\
			\bottomrule
		\end{tabular}
		\caption{Most frequent terms}
	\end{subtable}
	\caption{\label{tab:development_economics}Development Economics papers. A description of 1,564 Development Economics research articles. Table (a) lists the titles of the most cited papers in the subset. Table (b) lists terms that appear most frequently in the titles and abstracts of the papers, along with the most common MAG FoS. The number of articles with the respective terms/FoS are included in brackets.}
\end{table}